\documentclass[preprint,floatfix,amsmath,amssymb]{revtex4-1}
\usepackage{graphicx,float,amssymb,subfigure,dcolumn}
\begin{document}
	\title{Screening for Two dimensional MX$_2$ semiconductors with possible high room temperature mobility}
	\author{Zhishuo Huang, Wenxu Zhang
		\footnote{email to: xwzhang@uestc.edu.cn}, Wanli Zhang and Yanrong Li}
	\affiliation{State key laboratory of electronic thin films and integrated devices, University of
		Electronic Science and Technology of China, Chengdu 610054, China}
\begin{abstract}
	We calculated the electron mobility of 14 two dimensional semiconductors with composition of MX$_2$, where M (= Mo, W, Sn, Hf, Zr and Pt) is the transition metal, and X is S, Se and Te. We treated the scattering matrix by deformation potential approximation. Long wave longitudinal acoustical and optical phonon scatterings are included. Piezoelectric scattering in the compounds without inversion symmetry is also taken into account. We found that out of the 14 compounds, WSe$_2$, PtS$_2$ and PtSe$_2$, are promising regarding to the possible high electron mobility and finite band gap. The phonon limited mobility in PtSe$_2$ reaches about 3000 cm$^2$V$^{-1}$s$^{-1}$ at room temperature which is the highest among the compounds. The bandgap under the local density approximation is 1.25 eV. Our results can be a guide for experiments to search for better two-dimensional materials for future semiconductor devices.
\end{abstract}
\maketitle
\section{Introduction}
Two-dimensional (2D) layered materials have received a lot of research activities since the discovering  of graphene. New materials such as transitional metal dichalcogenides, especially MoS$_2$\cite{mos2}, and black phosphorus \cite{blackp} with one or several atomic layers were synthesized. Field effect transistors, photo detectors and  light emitting diodes were fabricated to demonstrate the potential applications of these materials. High speed radio frequency devices are fabricated which make full use of the ultrahigh electron mobility in graphene\cite{liao2010}. Carrier mobility is one of the key parameters for semiconductors to be used in the high speed or high frequency devices. Graphene has ultrahigh mobility, but it is intrinsically metallic. It is possible to open a gap at the Dirac cone, but usually the gap is tiny, and required extra parameters, for example, external electric fields\cite{castro07}, which is  not easy to fully integrate it with present semiconductor processes. When it was used in logical devices, the on/off ratio is  small, and the power consumption is large at the off-state. Monolayer MoS$_2$ has a direct band gap about 1.7 eV. However,  the electron mobility at room temperature is only up to ten cm$^2$V$^{-1}$s$^{-1}$. There are quite a lot of efforts to improve the carrier mobility. The mobility can be enhanced by even 20-fold to $\sim$200 cm$^2$V$^{-1}$s$^{-1}$ when encapsulated with a high-$k$ dielectric layer \cite{Radisavljevic2013}. This mobility enhancement was attributed to screening  of the charged impurity scattering by the dielectric layer\cite{ong}. On the device level, the mobility can be engineered by electron doping. WSe$_2$ monolayer with a direct band gap about 1.65 eV was explored after MoS$_2$\cite{fang}, which shows high field-effect mobility about 140 cm$^2$V$^{-1}$s$^{-1}$. It is already substantially higher than that of most of the 
reported room-temperature mobility values of MoS$_2$.  1H-MoTe$_2$ has a direct band gap a little above 1.0 eV. The room temperature electron mobility  was found to be about 30 cm$^2$V$^{-1}$s$^{-1}$ without gating\cite{lezama, pradhan}. This is already comparable with that of MoS$_2$.
\par According to the work of Yoon\cite{Yoon2011}, due to the heavier electron  effective mass and a lower mobility, MoS$_2$ is not ideal for high-performance semiconductor device applications. A natural question arises: can we find 2D materials with larger carrier mobility? And how far can we go to improve the mobility in the reported materials? Both of these questions can be answered by theoretical calculations. The ultramost mobility can be reached is limited by the intrinsic scattering. Phonon scattering is the one which cannot be avoided at temperatures other than zero. There are quite some works to calculate the phonon-limited carrier mobility in 2D materials.  The value in MoS$_2$ was first calculated by Kaasbjerg \emph{et al} \cite{kaasbjerg2012}. The contributions from acoustic and optical phonons are included and electron-phonon coupling matrices are calculated by the frozen phonon method. Electron-phonon, as well as piezoelectric interactions, are taken into account. The calculated room temperature mobility is about 410 cm$^2$V$^{-1}$s$^{-1}$ which sets the upper bound of electron mobility. This value estimated by Restrepo {\em et al.}\cite{restrepo} is 225 cm$^2$V$^{-1}$s$^{-1}$. Li\cite{Li2013} {\em et al.} developed the method to calculate the electron-phonon interactions based on the density functional perturbation theory and obtained similar results. The phonon limited electron mobility in single-layer MoS$_2$ is dorminated by the acoustic phonons at low field, while 
the degeneration by the optical phonons happens at high field according to the results of Monte Carlo simulations using the 
corresponding deformation potentials\cite{zeng}.  Very recently, Kim {\em et al.} \cite{kim14} calculated the phonon limited electron and hole mobility of four monolayer transition metal dichalcogenides: MoS$_2$, MoSe$_2$, WS$_2$ and WSe$_2$. WS$_2$ provides the best performances of electrons and holes with high mobilities. It was also found that p-type WSe$_2$ shows hole mobilities comparable or even larger than that of bulk silicon at room temperature. 
\par Our previous work shows that MoTe$_2$, ZrSe$_2$ and HfSe$_2$ have the largest electron mobility if only long wave acoustical phonon scattering was taken into account\cite{zhang}. However, as shown in the work of Kaasbjerg\cite{kaasbjerg2012}, the optical phonon may also play equally important role in MoS$_2$ in the room-temperature range due to the low optical phonon frequencies. In this work, we take these effects into account, and try to screen out MX$_2$ type 2D monolayer semiconductors with possible high electron mobility. We performed electronic calculations of the selected semiconductors with composition of MX$_2$, where M (= Mo W, Sn, Hf, Zr or Pt) is the transitional metal, and X is S, Se or Te. In order to fast screen out the materials with high performances, mobility limited by long wave phonons was estimated by calculating the corresponding deformation potentials. We found that out of the 14 compounds, WSe$_2$ with 1H-structure and PtSe$_2$ with 1T-structure, are promising regarding to their mobility and sizable band gap. We can see that due to the optical phonon scatter, only PtS$_2$ and PtSe$_2$ with the 1T-structure can have electron mobility larger than 1500 cm$^2$V$^{-1}$s$^{-1}$.
\section{Calculation details}
\par
\begin{figure}
	\includegraphics[scale=0.45]{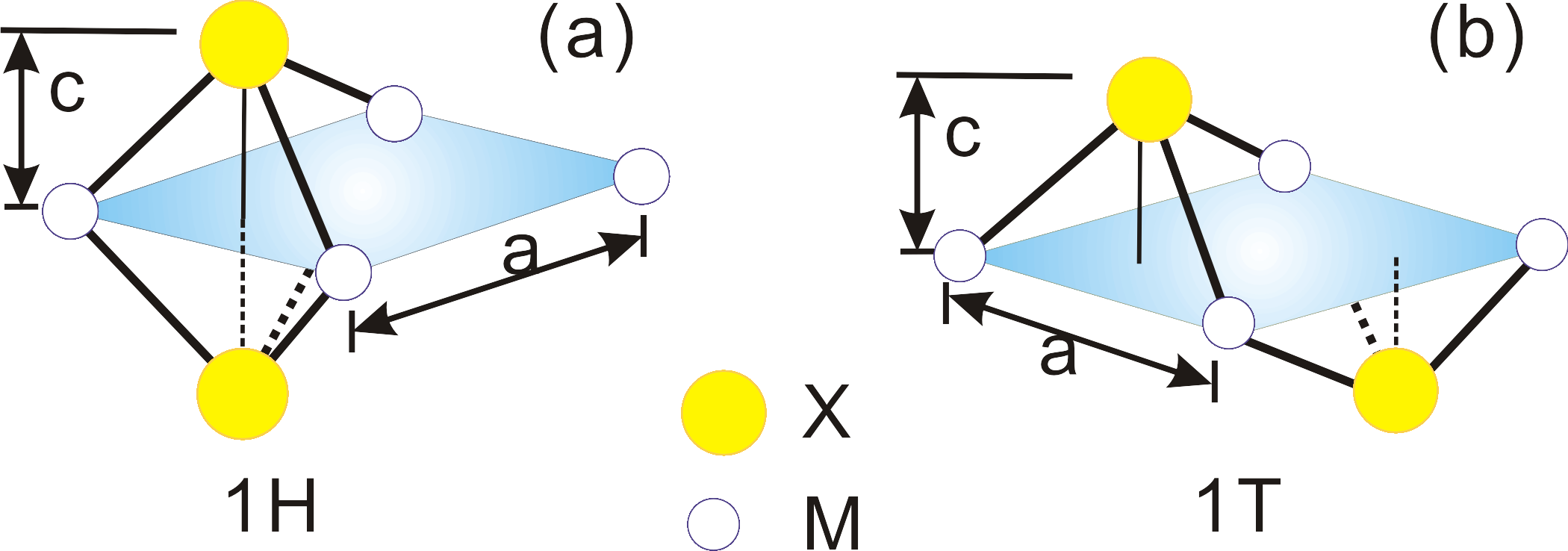}
	\caption{\label{fig:str} Schematic illustration of the atoms with the structure of MoS$_2$ (1H) and CdI$_2$ (1T).
		The lattice parameters are $a$ and $c$.}
\end{figure}

 Out of the ICSD database, only 16 compounds with MX$_2$ are semiconductors\cite{Lebegue2013}. There are more layered compounds with more chemical elements. The complexity may hinder their practical applications. We selected 14 MX$_2$ compounds which crystallizes into two different hexagonal crystal structures. One is the same as MoS$_2$ and the other is CdI$_2$. They are also called 1H and 1T structure, respectively. The difference is that the anion hexagonal nets are A-A stacked in MoS$_2$, while they are A-B stacked in CdI$_2$ as shown in Fig.\ref{fig:str}. Thus, the first one has a mirror plane of the metals and is lack of inversion, while the second one has the inversion center.

\par The calculations were performed mainly with pseudo-potential code PWscf\cite{QE}, and crosschecked by the full-potential local-orbital code (FPLO)\cite{fplo} in its current version with the default basis settings. All calculations were done within the scalar relativistic approximation. The local density approximation (LDA) functional was chosen to be that parameterized by Perdew and Wang\cite{lda}. The plane-wave kinetic energy cutoff was set to 90 Ry with density cut-off of 900 Ry. A shifted $17\times17\times2$ Monkhorst-Pack mesh was used to perform Brillouin zone integration in order to ensure the convergence of the results. Convergence of the total energy was set to be better than 10$^{-8}$ Hartree. A vacuum layer with thickness of 30 a.u. was used to model the 2D-nature  of the compounds. Forces on the atoms are limited within 0.001 Ry/a${_B}$ to obtain the ground states and total energy after full geometry relaxation. Phonon frequencies and phonon eigenvectors of at the $\Gamma$-point were calculated in a $4\times4\times1$ $q$-grid under the density functional perturbation theory (DFPT). The electron mobilities ($\mu$) were calculated by summation of the contributions of different scattering sources:
\begin{equation}
\frac{1}{\mu}=\frac{1}{\mu_{LA}}+\frac{1}{\mu_{OP}}+\frac{1}{\mu_{PZ}},\label{equ:mobility}
\end{equation}
where $\mu_{LA}$ is the longitudinal acoustic phonon determined mobility, $\mu_{OP}$ is the longitudinal optical phonon and $\mu_{PZ}$ is the piezoelectricity determined ones, respectively.
\section{Results and discussions}
\subsection{Electron and phonon dispersions}
\par The electronic structures have been discussed in our previous work\cite{zhang}. The most important parameters related here is the effective masses of electron. As seen in the previous works, the compounds with the 1T-structure are mostly  anisotropic, while the one with MoS$_2$ structure are more or less isotropic. The effective electron masses (m$^*$) and the LDA bandgap are listed in Table \ref{table:me}.  The m$^*$ of MoX$_2$ are about half of that of the free electron (m$_e$). The values of WX$_2$ are even lighter which are only about $\frac{1}{4}$ of m$_e$. With the increase of atomic number of X, it slightly increases.  \par
\begin{figure}
	\includegraphics[width=\textwidth]{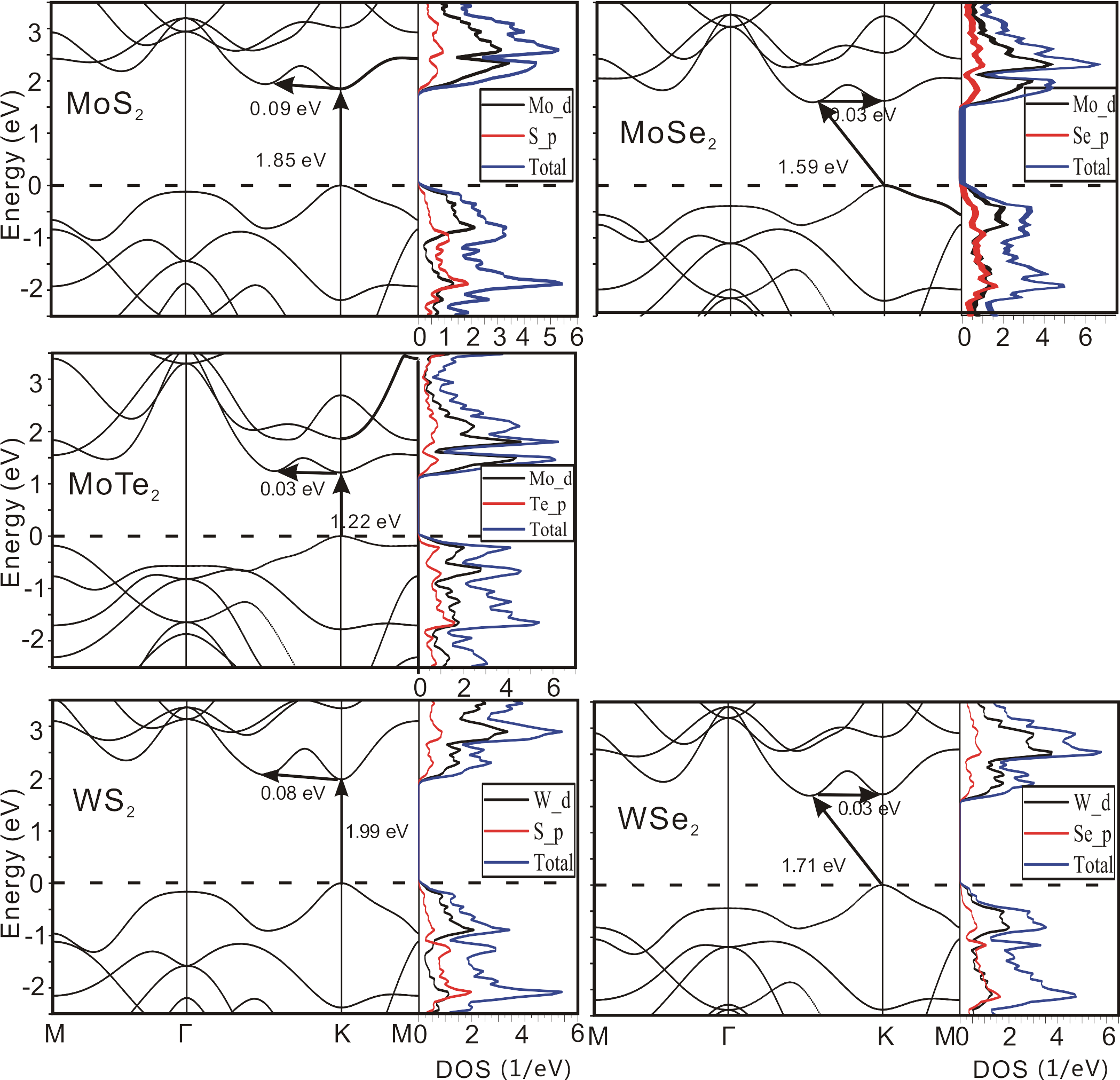}
	\caption{\label{fig:bands_H} (color online) Electronic bands of 1H structure compounds and the projected and total DOS.}
\end{figure}

\begin{figure}
	\includegraphics[width=0.75\textwidth]{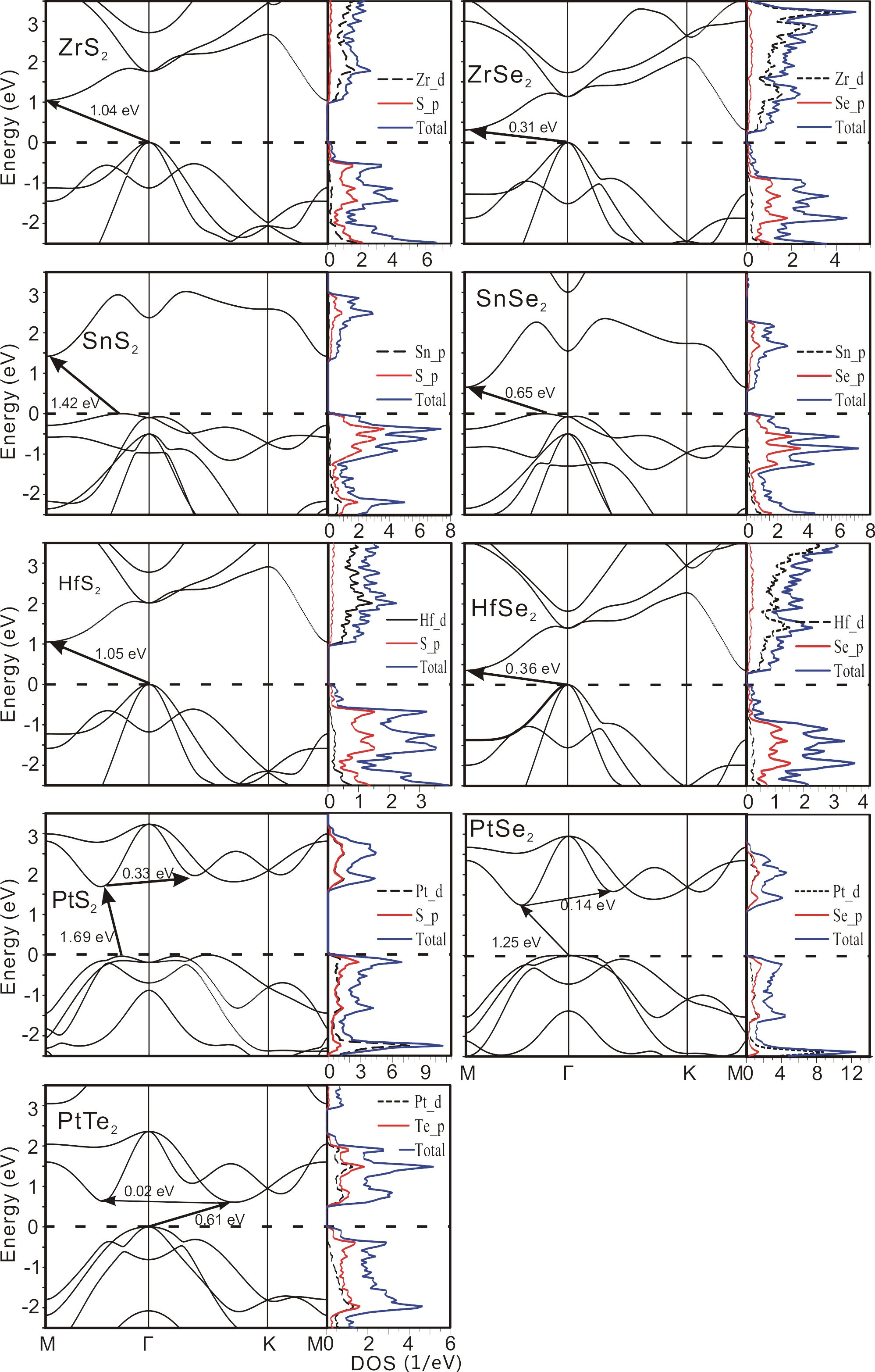}
	\caption{\label{fig:bands_T} (color online) Electronic bands of 1T structure compounds and the projected and total DOS.}
\end{figure}

 There is a sizable LDA bandgap in all of these compounds ranging from 0.31 eV in ZrSe$_2$ to 1.99 eV in WSe$_2$, although it is usually systematically underestimated by LDA. The electronic band structure and the atomic projected Density of States (pDOSs) are shown in Fig. \ref{fig:bands_H} and \ref{fig:bands_T} for the two structures. In the 1T-compounds, the conduction band minimum is located at the $M$-point except PtX$_2$. It lies between $\Gamma-M$ for X = S and Se, and $\Gamma-K$ for PtTe$_2$. 
\begin{table}
	\centering
	\caption{The lattice constant, effective mass and the bandgap of the compounds.  The effective mass in the $\Gamma-K$ direction for the MoS$_2$ structure and $\Gamma-M$ direction  for the CdI$_2$ structure is calculated.}\label{table:me}
	\begin{tabular}{lcccccc}
		\hline
		MX$_2$ &a    & c     & m$^\ast_{\Gamma-K(M)}$ & m$^\ast_{K-M}$ &E$_{el-ph}$ &E$_g$\\
		&(a.u.)&(a.u.)  &(m$_e$)&(m$_e$)&(eV)&(eV)\\
		\hline
		MoS$_2$  & 5.927 & 2.962 & 0.45 & 0.45 & 3.90&1.85 \\
		MoSe$_2$ & 6.168 & 3.156 & 0.52 & 0.52 & 3.65&1.59 \\
		MoTe$_2$ & 6.618 & 3.411 & 0.53 & 0.57 & 0.92&1.22 \\
		WS$_2$   & 6.047 & 2.992 & 0.24 & 0.26 & 3.92&1.99 \\
		WSe$_2$  & 6.166 & 3.164 & 0.33 & 0.31 & 3.78&1.71 \\
		\hline
		SnS$_2$  & 6.879 & 2.797 &2.11 & 0.21 & 3.55&1.42 \\
		SnSe$_2$ & 7.165 & 2.999 &2.91 & 0.17 & 2.91&0.65 \\
		HfS$_2$  & 6.731 & 2.750 &3.30 & 0.24 & 1.31&1.05 \\
		HfSe$_2$ & 6.944 & 2.978 &3.10 & 0.18 & 1.08&0.36 \\
		ZrS$_2$  & 6.817 & 2.771 &1.62 & 0.31 & 1.52&1.04 \\
		ZrSe$_2$ & 7.007 & 3.008 &2.03 & 0.22 & 1.25&0.31 \\
		PtS$_2$  & 6.670 & 2.327 &0.26 & 0.25 & 3.63&1.69 \\
		PtSe$_2$ & 6.978 & 2.464 &0.21 & 0.19 & 2.86&1.25 \\
		PtTe$_2$ & 7.485 & 2.634 &0.90 & 0.77 & 1.73&0.61 \\
		\hline
	\end{tabular}
\end{table}
There is another local conduction band minimum (Q-valley) only tens of meV higher than the global conduction band minimum in the compounds with 1H-structure.  It gives additional electron scattering which will lower the electron mobility. This is also the case in PtTe$_2$ with the 1T-structure. The DOS is projected onto different atomic orbital orbitals as shown in the figure. The transition metal and the X-atoms both contribute to the states of the conduction bands in the 1H compounds. In 1T-structure, the situation is more complicated. In PtX$_2$, where the Pt and X-atoms contribute equally to the states. However, in HfX$_2$ and ZrX$_2$, the contributions from the X-atoms are tiny and it is mainly due to the metal atoms. They are reversed in SnX$_2$.
\par
\begin{figure}
	\includegraphics[width=0.75\textwidth]{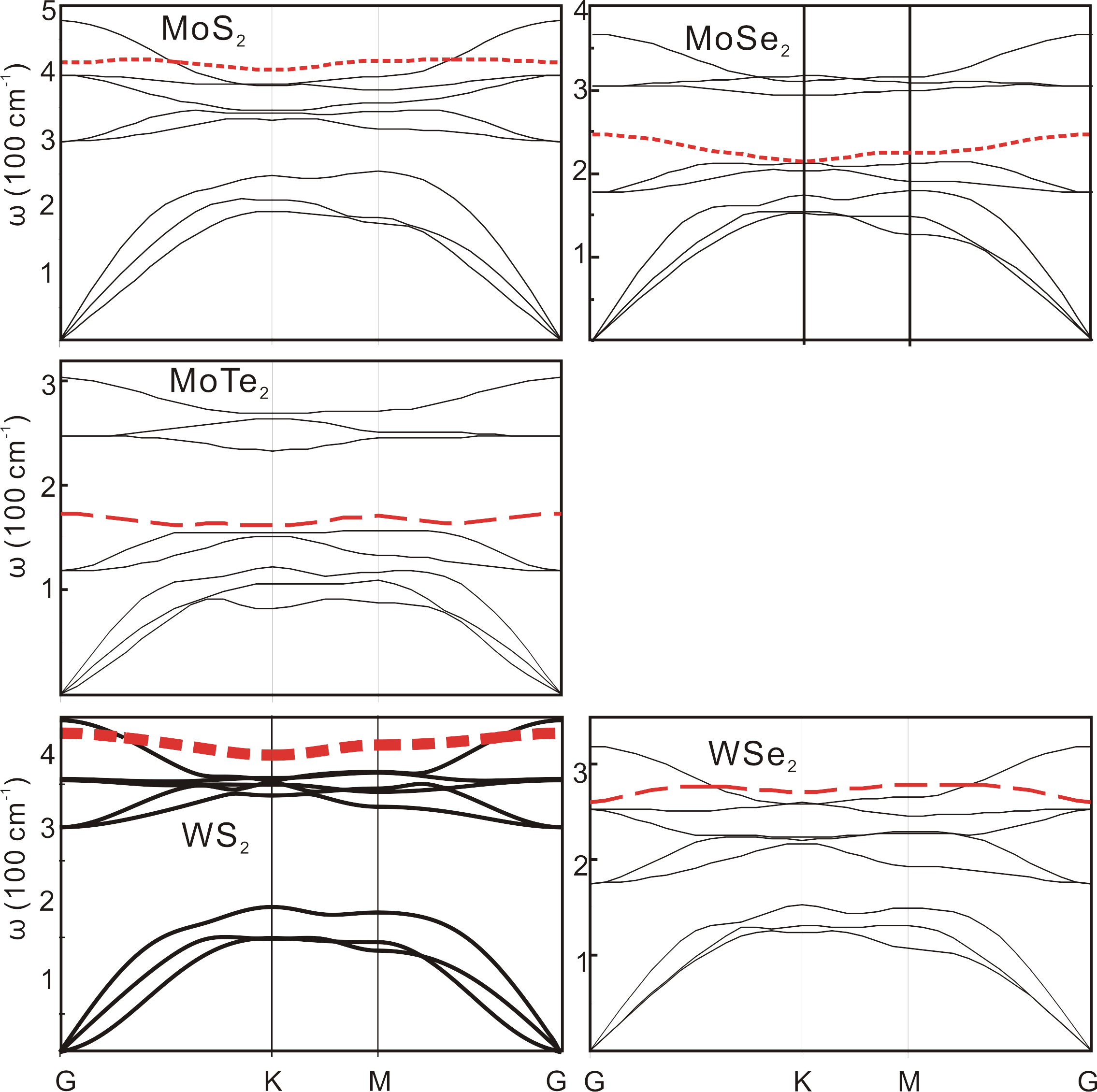}
	\caption{\label{fig:ph_H} (color online) The phonon dispersions of 1H structure compounds. The homopolar mode dispersion is in red.}
\end{figure}

\begin{figure}
	\includegraphics[width=0.6\textwidth]{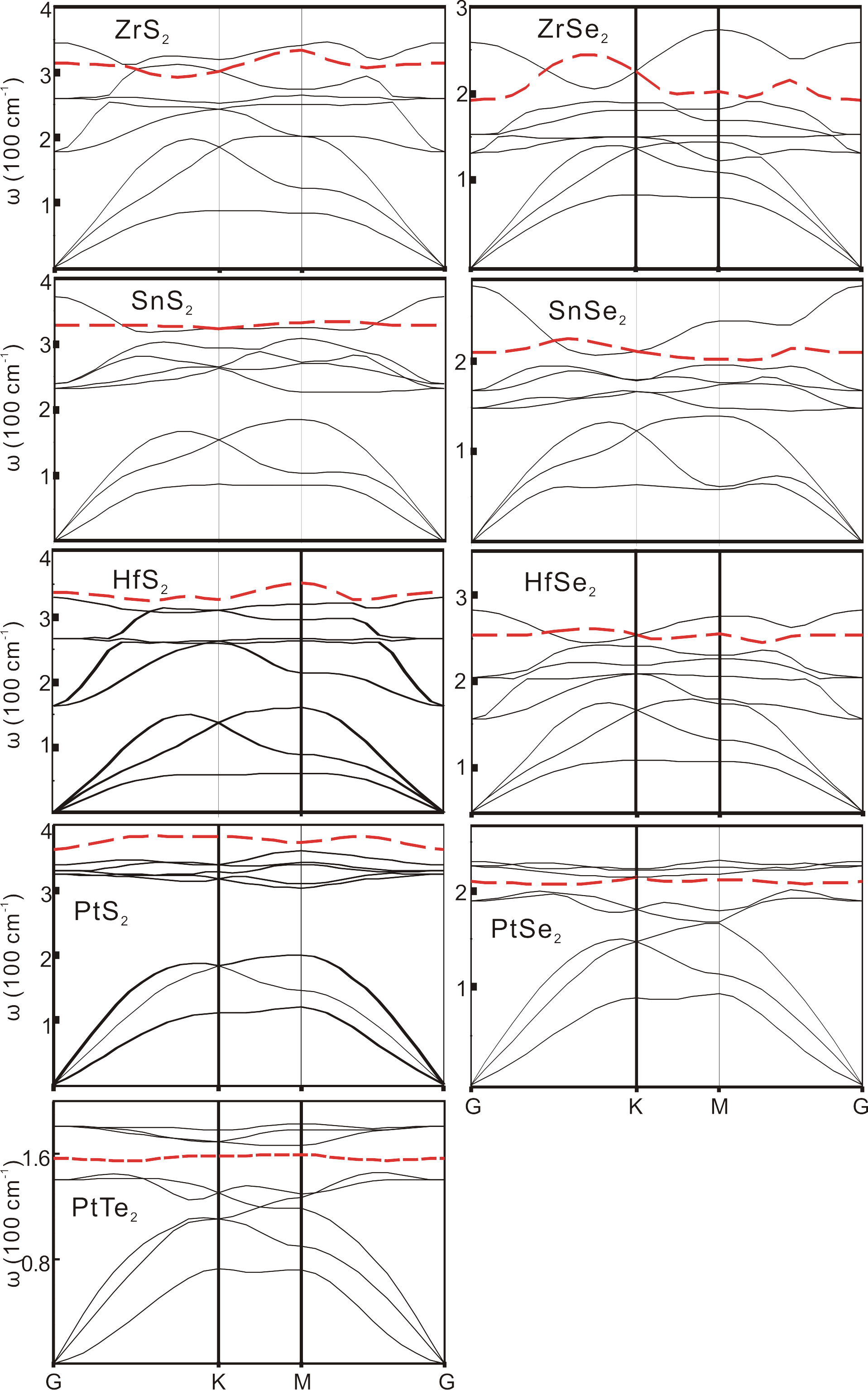}
	\caption{\label{fig:ph_T} (color online) The phonon dispersions of 1T structure compounds. The homopolar mode dispersion is in red.}
\end{figure}

 The phonon dispersions of these compounds were calculated by the DFPT method. The dispersions of the 1H and 1T structures are shown in Fig. \ref{fig:ph_H} and \ref{fig:ph_T}. The characteristic homopolar phonon of the two dimensional monolayer compounds are highlighted in the figure with red dashed lines.  Within the compounds with the same  metal atom, the frequency of the phonons increases with the atomic number. This is a consequence of the decrease of the chemical bond strength, where the forces between the atoms decrease, and the atomic weight increases. 
As the row number of X increases, the atoms become heavier, and the bonds become less ionic. Changes of phonon energy are systematic. Taken PtX$_2$ as an example, the width of the acoustical branches decease from 200 cm$^{-1}$ to 120 cm$^{-1}$. The highest optical phonon frequency deceases from 380 cm$^{-1}$ to 180 cm$^{-1}$. The gap between the optical and the acoustic branches also decreases with the decrease of the atomic weight differences. Especially, there is a gap about 100 cm$^{-1}$ between the optical and acoustic branches in PtS$_2$. However it diminishes in the other compounds with 1T-structure. The gaps also diminish in HfSe$_2$ and ZrX$_2$.
The larger phonon bandgap is favorite to produce larger thermal conductivity due to the reduced acoustical and optical phonon inter-scattering. As calculated by Gu {\em et al}\cite{gu}, the thermal conductivity of 1H-type MX$_2$ are above 50 W/mK at room temperature, while that values of the 1T-type are much lower. The flat homopolar branch are quite obvious as it is a characteristic of these layered structure compounds.
The linear dispersion of the TA branches becomes parabolic as the atomic number increases in PtX$_2$ due to the rapid decay of the transverse force constants.
\par Since the frequency of the optical phonon mode in these compounds is about tens of meV, which is comparable with  the thermal energy of 30 meV at T$=300$ K. It can have large populations at room temperatures. Thus both the acoustic and optical phonons may have the same order of contribution to the  scattering processes.  
\subsection{Acoustic phonon scattering}
The electron mobilities at $T=300$ K were approximated by the relaxiation time approximation. Within the deformation potential approximation\cite{Bardeen1950,takagi96}, the electron mobility (Takagi model) is approximated by
\begin{equation}
\mu_{LA}=\frac{e\hbar^3\rho V^2_s}{k_B T m^\ast m_d E^2_{el-ph}}.\label{equ:mob_la}
\end{equation}
where k$_B$ and $\hbar$ are the Boltzmann constant and the Planck constant, 
respectively. $m^\ast$ is the effective mass of electron in the electron propagation direction and $m_d$ is 
the electron density of state mass, which is $\sqrt{m_{\parallel}m_{\perp}}$, where $m_{\parallel(\perp)}$ 
is the mass parallel 
(perpendicular) to the propagation direction. $\rho$ is the mass density of the material and V$_s$ is the sound velocity in the corresponding direction. The validity of this model has already been demonstrated. Bruzzone and Fiori used this model to compute the electron mobility of hydrogenated and fluorinated graphene as well as h-BCN from first principles\cite{Bruzzone2011} and show that graphene with a reduced degree of hydrogenation can compete with silicon technology. The sound velocity (V$_s$) is calculated by the supercell method, where frozen phonon mode corresponding to the longitudinal phonon with vector $\mathbf{q}=\frac{\pi}{8a}(1,0,0)$ was simulated. The phonon frequency ($\omega_k$) was obtained, which is related to the sound velocity by $\omega_k=V_s |\mathbf{q}|$. 
The electron phonon coupling (E$_{el-ph}$) was approximated by the deformation potential D$_{ac}$. It is related to the variation of 
the electron eigenvalue, which is caused by the volume changes by
\begin{equation}
\Delta E_k=D_{ac}\frac{\Delta V}{V}.\label{equ:dac}
\end{equation}
The related values are listed in Table \ref{table:def}. The values of D$_{ac}$ calculated from fitting the relaxion time obtained by Kim\cite{kim14} and Kaasbjerg\cite{kaasbjerg2012} are also listed in the Table. They relative deviation is within 17 \%. However, the values of MoS$_2$ obtained by Kaasbjerg is 2.4 eV, compared with 4.5 eV by Kim and 3.9 eV by us, we can see that this value is method sensitive. But the scale in all these calculations is the same. 
\begin{table}
	\centering
	\caption{The sound velocity (V$_s$), the acoustical(D$_{ac}$) and the optical (D$_{op}$) deformation potential.}\label{table:def}
	\begin{tabular}{lcccc}
		\hline
		MX$_2$ & V$_s$   &D$_{ac}$ & D$_{op}$ \\
		& (km/s)  &(eV)     & (10$^8$ eV/cm)\\
		\hline
		MoS$_2$  & 7.93 & 3.90,2.4\cite{kim14},4.5\cite{kaasbjerg2012}&1.75,5.8\cite{kim14},4.1\cite{kaasbjerg2012} \\
		MoSe$_2$ & 6.01 & 3.65,3.4\cite{kim14}&1.10, 5.2\cite{kim14} \\
		MoTe$_2$ & 5.04 & 0.92&1.34 \\
		WS$_2$   & 6.67 & 3.92,3.2\cite{kim14}&2.34, 3.1\cite{kim14} \\
		WSe$_2$  & 5.55 & 3.78,3.2\cite{kim14}&1.12, 2.3\cite{kim14} \\
		\hline
		SnS$_2$  & 6.18 & 3.55&0.69 \\
		SnSe$_2$ & 4.83 & 2.91&0.38 \\
		HfS$_2$  & 5.86 & 1.31&0.99 \\
		HfSe$_2$ & 4.72 & 1.08&0.62 \\
		ZrS$_2$  & 7.21 & 1.52&1.12 \\
		ZrSe$_2$ & 5.42 & 1.25&0.75 \\
		PtS$_2$  & 6.61 & 3.63&1.06 \\
		PtSe$_2$ & 4.73 & 2.86&0.84 \\
		PtTe$_2$ & 4.89 & 1.73&0.95 \\
		\hline
	\end{tabular}
\end{table}

\subsection{Optical phonon scattering}
\begin{figure}
	\includegraphics[scale=0.45]{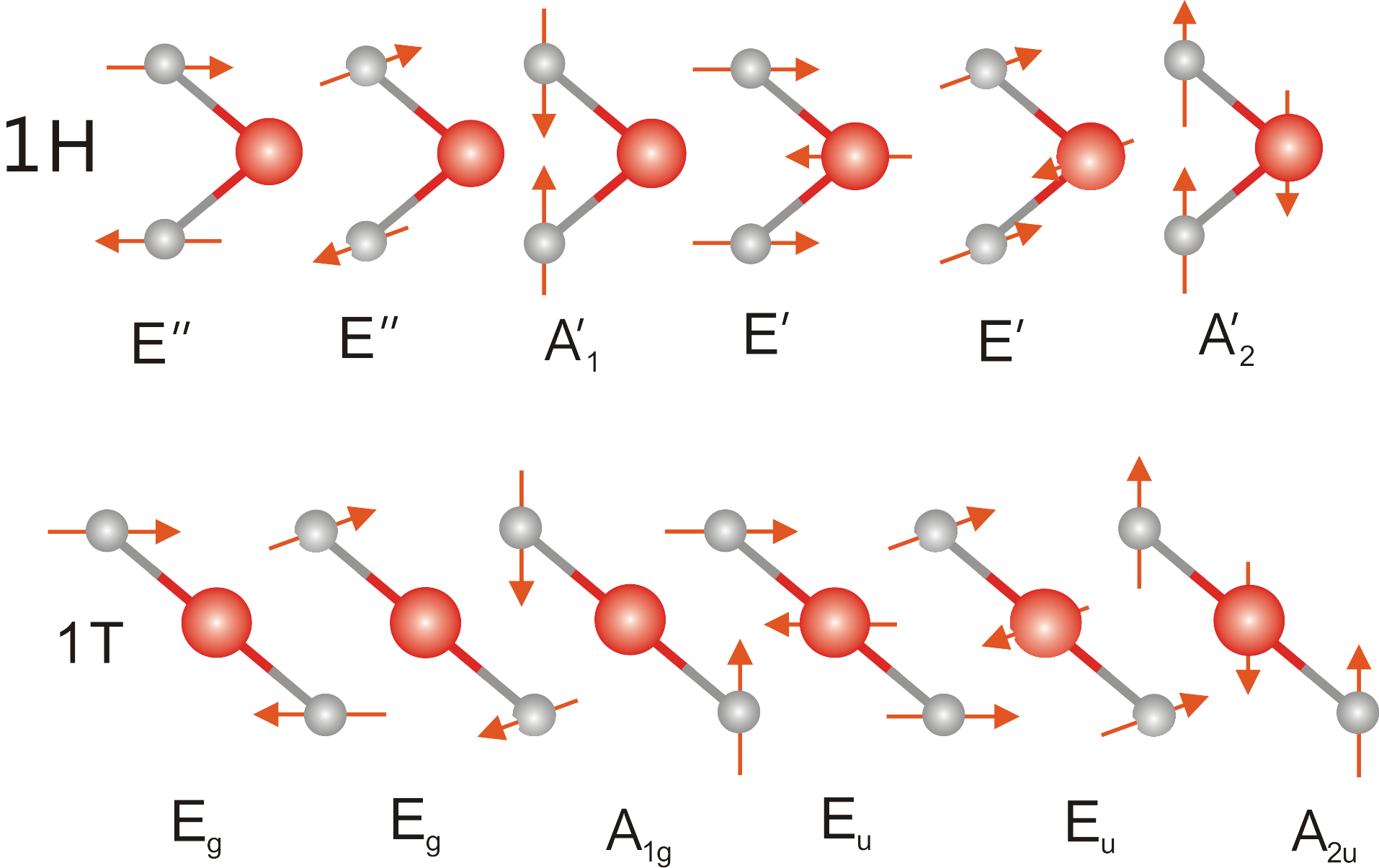}
	\caption{\label{fig:modes} The different optical phonon modes of  1H and 1T structures.}
\end{figure}

The zeroth order optical deformation potential D$_{op}$ is defined as 
\begin{equation}
\Delta E = D_{op}d
\end{equation}
where $\Delta E$ is the band shift under the proper deformation determined by the proper optical phonon mode.  $d$ is the atom displacement of the corresponding mode. There are six optical modes as shown in Fig. \ref{fig:modes}, corresponding to six deformation potentials of the conduction band minimum. 
The largest optical deformation potentials are listed in Table \ref{table:def}. The largest deformation potential was caused by the A$_{1g}$ (homopolar) mode of the phonon where X atoms move in the opposite direction. The only exceptions are PtS$_2$ and PtSe$_2$ where the contribution are from the $E{''}$ mode. As reported previous studies\cite{chakraborty}, the A$_{1g}$ mode is sensitive to electron doping and a red shift is observed on doping, which will increase the scattering rate. Thus we can expect a significant decrease of the mobility. However, the in-plane mode is less sensitive  to doping and the high mobility is expected to be kept. 
\par The optical phonon limited mobilities are calculated by Equ. (\ref{equ:opt}) according to the relaxation time approximation:
\begin{equation}
\mu_{OP}=\frac{2e\hbar^2\rho\omega_{\nu}}
{g_{d}m_{d}m^\ast D_{op}^{2}[N_{\nu}\Delta_{1}+(N_{\nu}+1)\Delta_{2}]}, \label{equ:opt}
\end{equation}
where g$_{d}$ is the valley degeneracy for the final electron states, N$_{\nu}$ is the occupation number of phonon with angular frequency $\omega_\nu$ of mode $\nu$ which is governed by Bose-Einstein distribution, and $\Delta_{1,2}$ are the onsets of scattering for phonon 
absorption and emission, both of which are set to 1 for intravalley optical phonon scattering. 
\subsection{Piezoelectric effects}
When the inversion center is missed in the polar semiconductors, optical deformation leads to electric polarization which will be a scattering source of carriers. The piezoelectricity induced scattering can be modeled in a similar way as the acoustic deformation  potential. The deformation potential $\Xi$ is replaced by \cite{kaasbjerg2013}
\begin{equation}
\Xi^{2} \rightarrow \frac{1}{2}(\frac{e_{11}e}{\epsilon_{r}\epsilon_{0}})^{2},
\end{equation}
where $\epsilon_{0(r)}$ is the vacuum (relative) permeability. The piezoelectric coefficient $e_{11}$ is defined  in 2D materials as linear response indicates piezoelectric effect could be treated as the first-order coupling between surface polarization and strain tensors ($\epsilon_{ij}$). The relation at fixed electric field $E$ and temperature $T$ is given by:
\begin{equation}
e_{ijk}=(\frac{\partial P_{i}}{\partial \varepsilon_{jk}})_{E, T}. \label{equ:pz}
\end{equation}
The symmetry required that they are related by
\begin{equation}
\begin{split}
e_{111}&=e_{11} \\
e_{122}&=e_{12}=-e_{11} \\
e_{212}&=e_{221}=e_{26}=-e_{11} \\
\end{split}
\end{equation}
With the above relations, the piezoelectric coefficient is calculated by
\begin{equation}
P_{1}(\varepsilon_{11},\varepsilon_{22}=0)-P_{1}(\varepsilon_{11}=0,\varepsilon_{22}=0)=e_{11}\varepsilon_{11}.
\end{equation}
In our calculation, the strain, ranging from -0.006 with steps of 0.002 to 0.006, is applied to the compounds. The calculated $e_{11}$'s are listed in Table \ref{table:pz} with comparison with the previous work\cite{kara}. 
\begin{table}
	\centering
	\caption{The piezoelectric constant ($e_{11}$), dielectric constants ($\epsilon_r$) and the equivalent piezoelectric 
		scattering potential $\Xi_{pz}$ }.\label{table:pz}
	\begin{tabular}{lcccc}
		\hline
		MX$_2$ &\multicolumn{2}{c}{$e_{11}$($10^{-10}$C/m) }&$\epsilon_r$ & $\Xi_{pz}$(eV) \\
		\cline{2-3}
		&this work&ref.\cite{kara}                      &             &     \\
		\hline
		MoS$_2$  & 2.98 &3.06& 4.26&5.59 \\
		MoSe$_2$ & 2.68 &2.80& 4.74&4.54  \\
		MoTe$_2$ & 2.57 &2.98& 5.76&3.56  \\
		WS$_2$   & 1.72 &2.20& 4.13&3.32  \\
		WSe$_2$  & 1.51 &1.93& 4.63&2.61 \\
		\hline
	\end{tabular}
\end{table}

It can be seen that the agreement is good, although our LDA values are systematically smaller than the GGA values reported by the work of Duerloo\cite{kara}. The $e_{11}$'s of the MoX$_2$ compounds are larger than the WX$_2$ compounds and decrease with the increase of the atomic number of X (X = S, Se, Te). The values are in the same order of bulk piezoelectric materials which makes significant contributions to the scattering of the electrons. However, this scattering can be screened by dielectric effects as shown in experiments. This might be the reason where the mobility of the electron can be tuned by dielectric gating. 

\subsection{The total mobilities}
\begin{figure}
	\includegraphics[scale=0.25]{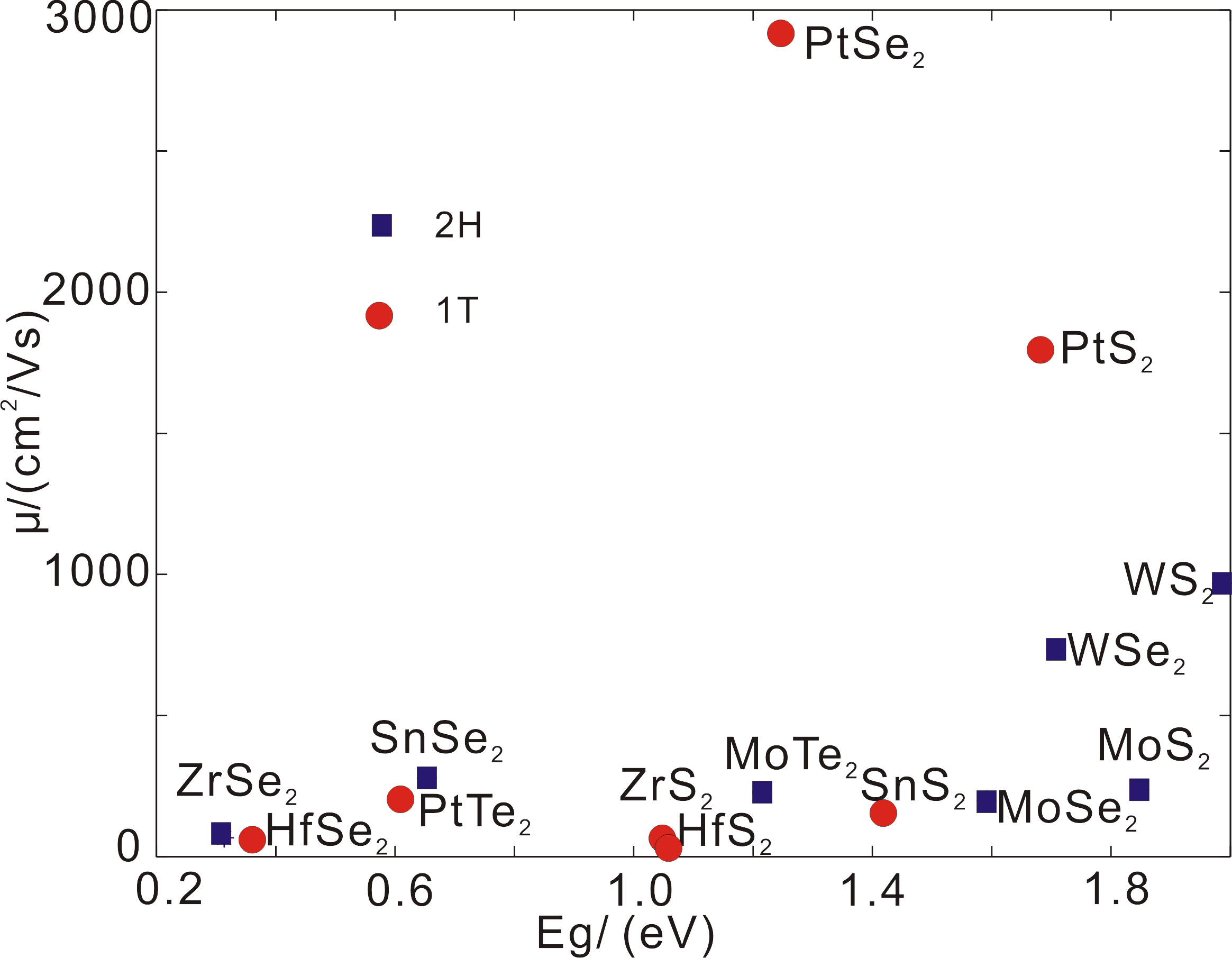}
	\caption{\label{fig:mu_eg} The total electron mobility and LDA band gap of the 14 compounds with 1H and 1T structure.}
\end{figure}

The total electron mobilities were calculated by the Matthiessen's rule as shown in Equ. (\ref{equ:mobility}), where the contributions from the longitudinal acoustic and optical phonons and the piezoelectric scattering were presented in the above sections. The data are listed in Table \ref{table:mu}. It can be seen that the optical phonon largely reduces the electron mobility, which is the main source of scattering. We can see that HfSe$_2$ and ZrSe$_2$ shown larger acoustic phonon limited mobilities, the optical phonon scattering are strong which reduces the electron mobility to be less than 100 cm$^2$V$^{-1}$s$^{-1}$. 
\begin{table}
	\centering
	\caption{The calculated electron mobilities (unit: cm$^2$V$^{-1}$s$^{-1}$) contributed from the different scattering sources.}\label{table:mu}
	\begin{tabular}{lcccc}
		\hline
		MX$_2$ &$\mu_{LA}$&$\mu_{OP}$&$\mu_{PZ}$&$\mu$\\
		\hline
		MoS$_2$& 1362 & 506 & 663 & 237\\
		MoSe$_2$& 963 & 420 & 621 &199 \\
		MoTe$_2$& 10104& 357& 673 & 228 \\
		WS$_2$& 4415 & 1553 & 6148 & 968 \\
		WSe$_2$& 2822 & 1191& 5921 & 734 \\
		\hline
		HfS$_2$& 7334 & 48 & - &47 \\
		HfSe$_2$& 14317& 59& - & 59 \\
		PtS$_2$& 4429 & 3027 & - & 1795 \\
		PtSe$_2$& 7568& 4746&-& 2916 \\
		PtTe$_2$& 1467 & 235 & - & 202 \\
		SnS$_2$& 1224 & 176 & - & 154 \\
		SnSe$_2$& 1788 & 329&- & 278 \\
		ZrS$_2$& 4989 & 65 & - & 64 \\
		ZrSe$_2$& 9264 & 67 &- & 67 \\
		\hline
	\end{tabular}
\end{table}
The total mobility and the LDA band gap are shown in Fig.\ref{fig:mu_eg}.
Among the compounds with the 1H structure, WS$_2$ and WSe$_2$ have mobilities of 968 and 734 cm$^2$V$^{-1}$s$^{-1}$, respectively. The LDA band gap is about 1.8 eV, which is only half of that of Si ($\sim$1750 cm$^2$V$^{-1}$s$^{-1}$). The well-studied MoS$_2$ has the upper limit of 237 cm$^2$V$^{-1}$s$^{-1}$ according to our calculation. The experimental value achieved now is about 200 cm$^2$V$^{-1}$s$^{-1}$, which is already quite near this theoretical upper limit. However, in the WS(Se)$_2$ compounds, the experimental values are only about tens of cm$^2$V$^{-1}$s$^{-1}$, which is far below our theoretical limit. The electron mobilities of Mo compounds are low, which is mainly caused by optical phonon scattering. This is in agreement with previous calculation and analyses by Kaasbjerg {\em et al.} 
\par In the compounds with 1T-structures, the piezoelectric scattering is lacking. Two compounds PtS$_2$ and PtSe$_2$ show the upper limits of electron mobilities of 1759 and 2916 cm$^2$V$^{-1}$s$^{-1}$, respectively. As can be seen from the above sections, the LDA band gaps are 1.69 and 1.25 eV. These compounds are thus promising when compared with Si. 
\par In this work, we have considered only the long wave longitudinal acoustic and optical phonon scattering. The initial and final states of the scattered electrons are limited to the bottom of the conduction band. That is to say, only the intravalley scattering is included. There are other scattering processes, such as interband scattering, and other scattering sources like impurities, electrons and so on. The mobility will be limited by any one of these mechanisms. There are uncertainties to precisely determine the contributions from each of these mechanism, both experimentally and theoretically, so that we cannot predict the mobilities precisely. However, by computing selected scattering sources, the upper bound of the mobility can be set. We thus can say, it is hopefully that we can find compounds with possible high mobility among the selected ones with larger upper bounds. It is NOT possible to find larger mobility in the compounds which are predicted to have lower theoretical upper bounds. More sophisticated calculations are needed to aim more precisely within the reduced number of compounds. However, these calculations are time consuming and not suitable for screening within the large amount of candidates.
\par According to our estimation, the mobility of WSe$_2$ and WS$_2$ may be larger than MoS$_2$. The mobility order is WS$_2>$ WSe$_2>$MoS$_2>$MoSe$_2$, which is the same as more elegant results by Kim {\em et al.}\cite{kim14}. Recently, mobilities of WSe$_2$ and MoS$_2$ are extracted from the transfer character curves of field-effect transistors \cite{Fang2013}. It is shown that the electron mobility in WSe$_2$ is about 110 cm$^2$V$^{-1}$s$^{-1}$, while that of MoS$_2$ is about 25 cm$^2$V$^{-1}$s$^{-1}$. These experimental results can be a preliminary confirmation of our prediction.
\section{Conclusions}
In this work, the electron mobility of 14 MX$_2$ type two dimensional semiconductors were calculated where only elastic scattering from long wave acoustic and optical phonons were taken into account by the deformation potential approximation. The piezoelectric scattering was included in compounds lack of inversion center and treated as the acoustic phonon-like deformation potential. We found that the total electron mobility in WS$_2$ can reach 968 cm$^2$V$^{-1}$s$^{-1}$, which is the highest value among the compounds with the 1H-structures. The value of 2916 cm$^2$V$^{-1}$s$^{-1}$ can be reached in PtSe$_2$ with the 1T-structure, where the inversion center is presented and the piezoelectric scattering is prohibited. WS$_2$ is a direct band gap semiconductor with band gap of 1.99 eV at the K-point, while PtSe$_2$ is an indirect band gap of 1.25 eV from the $\Gamma$-point to the middle of $\Gamma$-M line. Concerning with these two requirements, theses two compounds are promising for two-dimensional semiconductors used for future logical devices. 

\section{Acknowledgement}
Financial support from ``863''-project (2015AA034202), Research Grant of Chinese Central Universities (ZYGX2013Z001) and  NSFC (No.2011JTD0006) are acknowledged.

\end{document}